\documentclass[aps,twocolumn]{revtex4}
\usepackage{epsfig}
\newcommand{\be}{\begin{equation}}
\newcommand{\ee}{\end{equation}}
\newcommand{\ba}{\begin{eqnarray}}
\newcommand{\ea}{\end{eqnarray}}

\def\mtil{m_{\tilde{l}}}
\def\GeV{{\rm\ GeV}}

\begin{document}

\title{Propagation of Supersymmetric Charged Sleptons at High Energies}
\author{M. H. Reno}
\affiliation{%
Department of Physics and Astronomy, University of Iowa,
Iowa City, Iowa 52242 USA}
\author{I. Sarcevic and S. Su}
\affiliation{%
Department of Physics, University of Arizona, Tucson, Arizona 85721,
USA}
\begin{abstract}

The potential for neutrino telescopes to discover charged
stau production in neutrino-nucleon interactions in Earth
depends in part on the stau lifetime and range. In some 
supersymmetric scenarios, the next lightest supersymmetric
particle is a stau with a decay length on the scale of 10 km.
We evaluate the electromagnetic energy loss 
as a function of energy and stau mass. 
The energy loss parameter $\beta$ scales as the inverse stau mass
for the dominating electromagnetic processes,
photonuclear and $e^+e^-$ pair production.
The range can be parameterized as a function of stau mass,
initial energy and minimum final energy. In comparison to
earlier estimates of the stau range, our results are as much 
as a factor of two larger, improving the potential for stau 
discovery in neutrino telescopes.

\end{abstract}

\maketitle

\section{INTRODUCTION}

Neutrino telescopes have great potential for probing 
physics beyond the Standard Model 
when new particles are created by high energy cosmic 
neutrino passing through the Earth. 
A recent proposal \cite{chacko} showed that the detection of 
two nearly parallel charged particle tracks from a pair
of metastable staus,  
which are produced as secondary particles 
in very high energy neutrino-nucleon interactions, 
may provide a unique way of probing the supersymmetry (SUSY) 
breaking scale in weak scale supersymmetry models.

Ultrahigh energy neutrinos from astrophysical sources travel undeflected
to the Earth.  They would then interact with the nucleons in the Earth
and pair produce two superparticles via the 
$t$-channel chargino or neutralino exchanges: 
$\nu N \rightarrow \tilde{l} \tilde{q}$.  
In SUSY models where gravitino is the lightest supersymmetric particle (LSP)
and stau is the  next lightest supersymmetric particle (NLSP),
heavier slepton and squarks 
instantly decay into the stau NLSP \cite{susyref}.  
The high energy stau would then travel a long 
distance, typically more than tens to $10^4$ kilometers, 
before it decays into the gravitino LSP.
The superparticle pair production rate, however, 
is very small in comparison to the standard model charged-current 
cross section.  Thus
production of staus from downward neutrinos is negligible compared to
showers from production through the standard model processes.  However,
the observability of the stau produced via 
upward going neutrinos could be enhanced.  
The small stau production cross section are compensated by the enlarged 
effective volume of an instrumented region 
due to the very long lifetime of the stau and 
its relatively slow energy loss 
in the water, ice or rock surrounding the detector \cite{chacko}.

The potential for neutrino telescopes to discover stau 
production in neutrino-nucleon interactions in Earth
depends in part on the stau lifetime and range.
In order to refine estimates of the signal from metastable staus, 
it is crucial to determine the 
energy loss of the high energy stau as it
traverses the Earth.  
The average energy loss of a particle traversing a distance 
$X$  is given by 
\begin{equation}
-\langle{dE\over dX}\rangle = \alpha +\beta E
\label{eq:dedx}
\end{equation}
where $E$ is the energy of the particle, $\alpha$ describes 
ionization energy loss,
and $\beta$ is the radiative energy loss.  
Ref.~\cite{chacko} estimated the stau range 
using Eq. (\ref{eq:dedx}) with 
the approximation that both $\alpha$ and
$\beta$ are constants with respect to energy.  
In this paper, we evaluate both the energy and mass dependence of the 
stau electromagnetic energy loss. Our results are
applicable to any metastable heavy charged slepton. We determine 
$\beta$ and we evaluate the average stau range using
a one-dimensional Monte Carlo simulation of stau propagation
in rock. We comment on the
implications of the new results
for the evaluation of event rates by Albuquerque, Burdman and 
Chacko in Ref. \cite{chacko}.

This paper is organized as follows.  In Sec.~\ref{sec:energyloss}, 
we analyzed in detail the stau energy loss through 
photonuclear scattering, pair production and bremsstrahlung. 
In Sec.~\ref{sec:range}, we show the stau range using both a
one-dimensional Monte Carlo evaluation and a parametrized 
approximate formula.  In Sec.~\ref{sec:discussion}, we conclude.

\section{Stau Energy Loss} 
\label{sec:energyloss}
The radiative energy loss $\beta$ receives contributions from 
bremsstrahlung, pair production and photonuclear scattering: 
\begin{equation}
\beta^i (E) = {N\over A}\int_{y_{min}}^{y_{max}} 
dy\  y{d\sigma^i(y,E)\over dy},
\end{equation}
where $y$ is the fraction of stau energy loss in the radiative interaction:
\begin{equation}
y={E-E'\over E}
\label{eq:ydef}
\end{equation}
for final stau energy $E'$. The superscript $i$ denotes
bremsstrahlung (brem), pair production (pair) and photonuclear (nuc)
processes. Avogadro's number is $N$ and the atomic mass number of the target
nucleus is $A$.

\begin{figure}[t]
\begin{center}
\epsfig{file=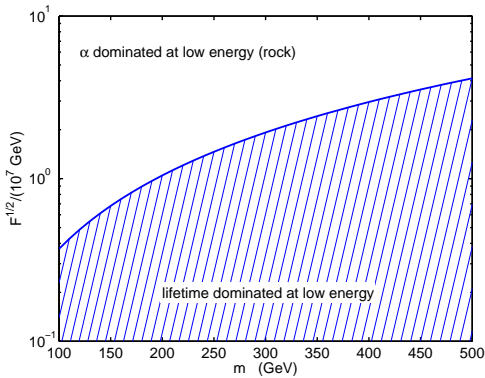,width=2.55in,angle=0}
\end{center}
\caption{Parameter space in $(m_{\tilde{l}},\sqrt{F})$ in which the
stau range is dominated by either ionization energy loss or the
lifetime, when $\beta E \ll \alpha $.}
\label{fig:alphatau}
\end{figure}

As we show below, the value for $\beta$ decreases with
increasing stau mass. The
ionization
energy loss parameter $\alpha$ is nearly constant as a function of
mass of the particle, namely \cite{ionization}
\begin{equation}
\alpha\simeq 2\times 10^{-3} \ {\rm GeV\, cm^2/g}\ .
\end{equation}
For low enough energies, where $\beta E\ll \alpha$ 
(for $\beta\sim 5\times 10^{-9}$
cm$^2$/g$\cdot (150\ {\rm GeV}/\mtil)$, it
corresponds to $E\ll 4\times 10^5$ GeV$\cdot(\mtil/150\ {\rm GeV})$), 
either the lifetime
or ionization energy loss determines the stau range,
which scales linearly with energy. This is shown in Fig.~\ref{fig:alphatau}. 
These two regions are separated by the line which corresponds to
$c\tau\rho=\mtil/ \alpha$, 
where $\mtil$ is the stau mass, $\rho $ is the density, taken
to be the density of standard rock $\rho=2.65$ g/cm$^3$, and
\begin{equation}
c\tau = \Biggl( \frac{\sqrt{F}}{10^7\ {\rm GeV}} \Biggr)^4
\Biggl( \frac{100\ {\rm GeV}}{\mtil}\Biggr)^5\ 10\ {\rm km}
\end{equation}
is the stau lifetime. The quantity $\sqrt{F}$ is the supersymmetry
breaking scale.  In low energy SUSY breaking models, the SUSY breaking 
scale $\sqrt{F}$ could be as low as $10^7$ GeV\cite{susyref}, 
leading to a stau  decay length of 10 km or longer.

\begin{figure}[t]
\begin{center}
\epsfig{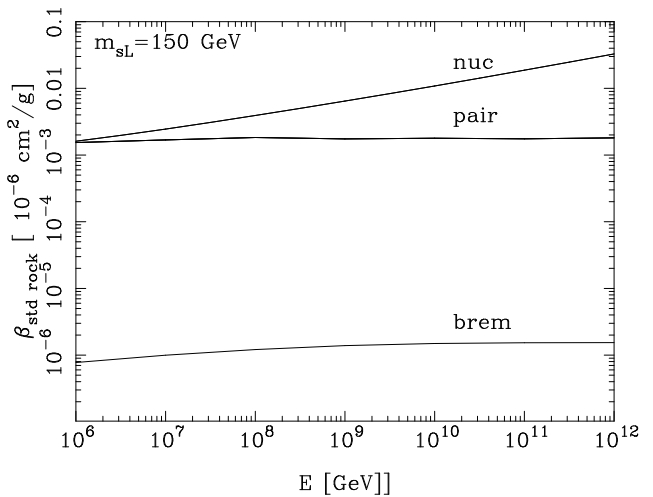}
\end{center}
\caption{Energy loss of stau due to photonuclear (nuc) 
interactions, pair production (pair)
and bremsstrahlung (brem) for a stau mass of 150 GeV.}
\label{fig:beta1}
\end{figure}

\begin{figure}[t]
\begin{center}
\epsfig{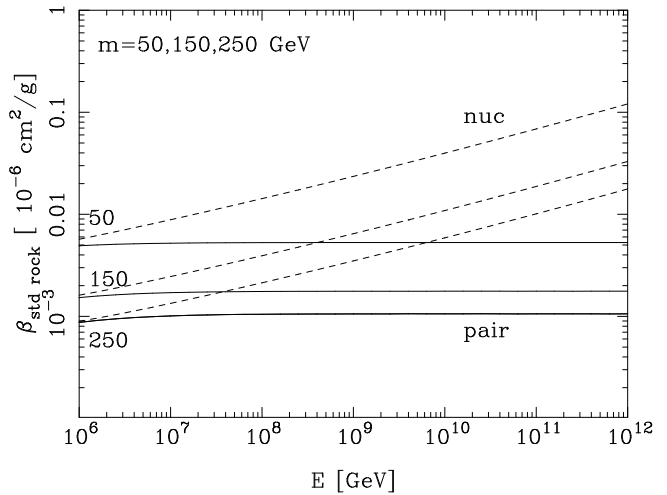}
\end{center}
\caption{Energy loss of stau due to photonuclear and pair production for 
$m_{\tilde \tau}=50$ GeV, 
$m_{\tilde \tau}=150$ GeV and 
$m_{\tilde \tau}=250$ GeV.} 
\label{fig:beta2}
\end{figure}

\subsection{Stau Energy Loss due to Photonuclear Interactions}

We obtain a photonuclear cross section that 
describes the interactions of charged
staus ($\tilde{\tau}$, or more generically charged sleptons
$\tilde l$) with nuclei via 
virtual photon exchange. 
Our approach \cite{drss} 
is to treat the $\tilde l N\rightarrow \tilde l X$ using the
deep-inelastic scattering formalism \cite{cs} in terms of the
two nucleon electromagnetic structure functions $F_1$ and
$F_2$.

We use $F_2$ parameterized by  Abramowitz, Levin, Levy
and Maor (ALLM) \cite{allm} to fit HERA data for a wide range
of $Q^2$ and $x$. The ALLM parameterization of the structure function
is based on the Regge approach and consists of a Pomeron and
a Reggeon contribution \cite{allm}. 
Other approaches \cite{bugaev}, which focus on the diffractive
region, but also include a perturbative component, 
give qualitatively similar results
for the photonuclear energy loss of leptons.

For stau scattering with nucleons, 
$\tilde{\tau}(k) N(p)\rightarrow \tilde{\tau}(k^\prime) X$, 
the differential cross section is given by
\begin{eqnarray}
\frac{d^2\sigma (x,Q^2)}{dQ^2\, dx} &=& \frac{4\pi \alpha^2}{Q^4}
\frac{F_2(x,Q^2)}{x} \Biggl[ 1-y+\frac{y^2}{4} \\ \nonumber
&&-\Biggl( 1+\frac{4\mtil^2}{Q^2} \Biggr)
\frac{y^2(1+4M^2x^2/Q^2)}{4(1+R(x,Q^2))}\Biggr]\ ,
\end{eqnarray}
where
$-Q^2= q^2=(k-k')^2$, 
$x={Q^2/( 2p\cdot q)}$, 
$y$ as in Eq. (\ref{eq:ydef}), 
$m_{\tilde l}$ is the stau mass, and $M$ is the nucleon mass.
$R(x,Q^2)$ parametrizes the violation of the Callan-Gross relation
$2xF_1(x,Q^2)=F_2(x,Q^2)$: 
\begin{equation}
R(x,Q^2) = \frac{\bigl( 1+{4M^2x^2\over Q^2}\bigr) F_2(x,Q^2)
-2x F_1(x,Q^2)}{2xF_1(x,Q^2)} \ .
\end{equation}
Since $R(x,Q^2)$ is small \cite{rstasto}, we approximate $R=0$.

To obtain $\beta$, we
convert the differential cross section to a dependence on
$y$ and $Q^2$: 
\begin{equation}
\frac{d^2\sigma (y,Q^2)}{dQ^2\, dy} = \frac{x}{y}\frac{d^2 \sigma
(x,Q^2)}{dQ^2\,dx}\ .
\label{eq:chvar}
\end{equation}
We use the following limits of integration:
\begin{eqnarray}
Q^2_{\rm min}&\leq& Q^2\leq 2MEy - ((M+m_\pi)^2-M^2)\\
y_{\rm min} & \leq & y\leq 1-\mtil/E\ ,
\end{eqnarray}
where $Q^2_{\rm min}\simeq \mtil^2y^2/(1-y)$ and $y_{\rm min}\simeq
((M+m_\pi)^2-M^2)/(2ME)$.

For comparison, the photonuclear cross section for the standard model
partner, the charged lepton, is given by \cite{drss}
\begin{eqnarray}
\frac{d^2\sigma (x,Q^2)}{dQ^2\, dx} &=& \frac{4\pi \alpha^2}{Q^4}
\frac{F_2(x,Q^2)}{x} \Biggl[ 1-y -\frac{Mxy}{2E} \\ \nonumber
&&+\Biggl( 1-\frac{2m_l^2}{Q^2} \Biggr)
\frac{y^2(1+4M^2x^2/Q^2)}{2}\Biggr]
\end{eqnarray}
in the $R\simeq 0$ limit. At high energies, the scalar and fermion
evaluation of $\beta^{\rm nuc}$ give the same numerical values for
equal scalar and fermion masses because the energy loss is dominated
by the small $y$ region of integration.

Our result for $\beta^{\rm nuc}$ for $\tilde \tau$'s with a mass of
150 GeV 
is given by the upper curve in Fig.~\ref{fig:beta1}.  
For $m_{\tilde{\tau}}=150$
GeV,
it is increasing with energy from $\sim 2\times 
10^{-9}$ cm$^2$/g at $10^6$ GeV to
$3\times 10^{-8}$ cm$^2$/g at $10^{12}$ GeV.

The dashed curves in Fig.~\ref{fig:beta2} 
represent the photonuclear contribution
to $\beta$ for three stau masses: 50 GeV, 150 GeV and 250 GeV, showing
that $\beta$ has a $1/\mtil$ dependence on the stau mass. This can
be understood based on the argument which has been discussed in the 
context of electron-positron production in muon
scattering by Tannenbaum in Ref. 
\cite{tannenbaum}. Qualitatively,
\begin{equation}
\label{eq:fapprox}
F_2\sim \frac{Q^2}{Q^2+m_0^2} f_2
\end{equation}
where to first approximation, $f_2$ is constant for small $y$. The
parameter
$m_0^2$ is 0.32 GeV$^2$ for the Abramowitz et al. parameterization
\cite{allm}. We
can approximate
\begin{equation}
\beta\simeq \int dy \int_{Q^2_{\rm min}}^{Q^2_{\rm max}} dQ^2\,
\frac{4\pi\alpha^2}{Q^2(Q^2+m_0^2)} f_2\ .
\end{equation}
The integral over $Q^2$ is approximately independent of $Q_{\rm max}^2$
and depends on $Q^2_{\rm min}/m_0^2$. In the small $y$ limit,
$Q^2_{\rm min}\simeq \mtil^2 y^2$. The change of variables of the
integral
from $y$ to $z$ where
\begin{equation}
z=\frac{\mtil y}{m_0} 
\end{equation}
yields a Jacobian factor of $m_0/\mtil$ multiplying the integral.
The stau mass dependence of $\beta$ is given by this Jacobian factor.

\subsection{Stau Energy Loss due to Pair Production}

Next we consider $e^+e^-$ pair production in collisions of staus with nuclei.
In this section, we show that the dominant contribution to $\beta$
for the production of an $e^+ e^- $ pair by a scalar charged particle
interaction with a bare nucleus is the same, at high energies, as for
the fermion case. We then use the results of Kokoulin and
Petrukhin for fermion scattering
\cite{koko} which include all of the diagrams and
incorporate a smooth transition
between the unscreened to screened nucleus.

\begin{figure}
\resizebox{3.5 in}{!}{\includegraphics*[60,50][550,720]{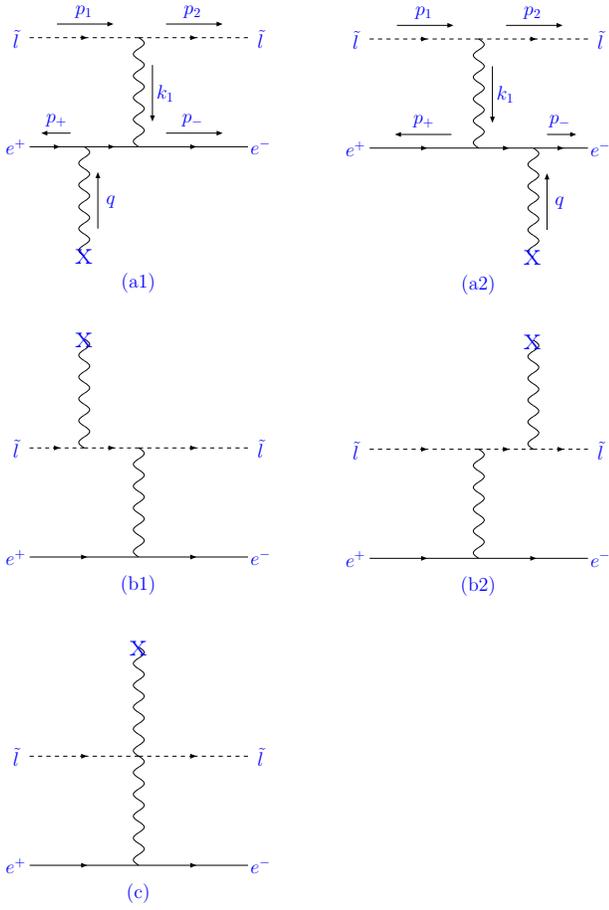}}
\caption{Feynman diagrams for $e^+ e^-$ pair production
by stau scattering from a nucleus with nuclear momentum
transfer $q$. Diagrams (a1) and (a2) dominate
$\beta^{\rm pair}$.}
\label{fey1}
%\end{center}
\end{figure}

This process is a very important contribution to the energy loss of muons 
and taus.  In the case of staus, it is also important. 
There are several diagrams that 
contribute.  They are shown in Fig.~\ref{fey1}.  

The diagrams can be classified into two
categories. Diagrams (a1) and (a2) are ``two photon'' contributions
to $e^+e^-$ pair production. Diagrams (b1), (b2) and (c)
combine to make the gauge invariant set of ``virtual bremsstrahlung''
contributions to $e^+e^-$ pair production. For taus instead of
staus, the virtual bremsstrahlung graphs are of the type (b1)
and (b2). The interference term between two photon (charge conjugation
even)  and virtual
bremsstrahlung (C-odd) contributions vanishes in the integral.

At relativistic energies, the two photon contribution
dominates because of the presence of two quasi-real photons.
This is independent of the spin of the charged particle
(tau or stau) that emits one quasi-real photon. 
The two photon process leads to a factor proportional to the square
of the classical electron radius $r_e=e^2/m_e c^2$, while the virtual
bremsstrahlung contribution is proportional to the square
of the classical heavy
lepton radius \cite{blp}, a result that carries over to the heavy
slepton case at high energy.
The dominance
of the two photon process was
first demonstrated in Ref. \cite{ll34} and has
been studied in many contexts, including lepton pair production
in colliding beam experiments \cite{brodsky}, and more recently 
in heavy ion collisions \cite{nuclth}.

Since the two photon
diagrams dominate, we calculate only these two contributions.
The cross section corresponding to 
these diagrams, in the notation of Kelner\cite{kelner}, 
is given by:

\begin{eqnarray}
\nonumber
{d\sigma} &=& \frac{
Z^2\alpha^4}{\pi^4} 
[{(p_1\cdot u)}^2-\mtil^2]^{-1/2} \frac{1}{k_1^4 Q^4} A^{\alpha\beta} 
B_{\alpha\beta}^{\mu\nu} 
u_\mu u_\nu  \\  && \cdot
\frac{d\mathbf{p}_+}{p_+^0} 
\frac{d\mathbf{p_-}}{p_-^0} 
\frac{d\mathbf{p_2}}{p_2^0} 
\frac{d\mathbf{q}}{u_0} 
\delta ( q + k_1 - p_+ - p_-) 
\end{eqnarray} 
where 
$u_\mu$ is the velocity of the nucleus, and $p_1$, $p_2$, $p_+$ and 
$p_-$ are the momentum for incoming stau, outgoing stau, 
(outgoing) position and electron, respectively.
In addition, $k_1=p_1-p_2$, $q=p_+ +p_- -k_1$, and $Q^2=-q^2$.
For scalar
interactions,
\begin{equation}
A^{\alpha\beta} = (p_1+p_2)^\alpha (p_1+p_2)^\beta \ .
\end{equation}
The quantity $B_{\alpha\beta}^{\mu\nu}$ is the usual fermion
trace involving the sum of the two diagrams with the two
photon vertices. A series of separations and integrals can
be performed so that the differential cross section can be
written in the following form \cite{ivanov}:
\begin{eqnarray}
d\sigma &=& \frac{2}{3}\frac{Z^2\alpha^4}{\pi m_e^2} I_1 (1-y)
dx_+\frac{d\omega}{\omega} dQ^2\\ \nonumber
&& \cdot G(x_+,x_-,Q^2,y)\ ,
\end{eqnarray} 
where
\begin{eqnarray*}
x_+&=&\frac{p_+^0}{\omega},\quad x_-=\frac{p_-^0}{\omega}\\
\omega &=& p_+^0 + p_-^0\\
y&=&\frac{\omega}{p_1^0}\\
I_1&=&\log \Biggl[ \frac{4\omega ^2 x_+^2
  x_-^2}{m_e^2\gamma}\Biggr]-1\\
\gamma &=& 1+\frac{Q^2}{m_e^2}x_+ x_-\ .
\end{eqnarray*}
One finds that $G=G(x_+,x_-,Q^2,y)$ is
\begin{eqnarray}
\label{eq:gval}
G&=& a_1\Bigl( \frac{1}{Q^2}-\frac{x_+x_-}{\gamma
  m_e^2}
\Bigr) - b_1\frac{\xi m_e^2}{Q^4 x_+ x_-}\nonumber \\
&& -c_1 \frac{x_+ x_-}{m_e^2 \gamma^2}\\ \nonumber
a_1 &=& (2 x_+^2+2 x_-^2+1)(D+\lambda C)\\ \nonumber
&& + 2D\xi (x_+^2+x_-^2+1)\\ \nonumber
b_1 &=& D(2x_+^2+2x_-^2+1) \\ \nonumber
c_1 &=& D(1+\xi )+\lambda C \\ \nonumber
&& + 2x_+x_-[\lambda C-(2+\lambda)D]
\end{eqnarray}
with the additional definitions
\begin{eqnarray}
\xi &=& \frac{\mtil^2}{m_e^2}x_+x_- 2\lambda\\
\lambda &=& \frac{y^2}{2(1-y)}\ .
\end{eqnarray}
The quantities 
$C$ and $D$ \cite{ivanov} characterize the incident particle
as scalar or fermion:
\begin{eqnarray*}
C=0 &\quad& D=1\quad\quad{\rm scalar\ case}\ , \\
C=1 &\quad& D=1\quad\quad{\rm fermion\ case}\ .
\end{eqnarray*}

\begin{figure}[t]
\epsfig{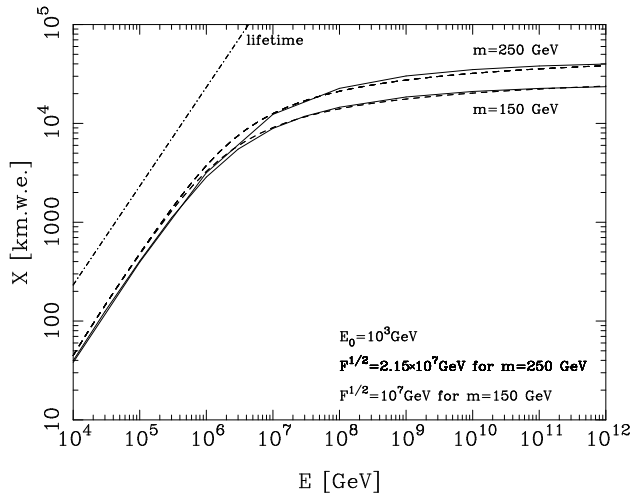}
\caption{Stau range (solid) and time dilated decay length
(dot-dashed line) for different values of $\sqrt{F}$ and for
$\mtil=150$ GeV and $\mtil=250$ GeV. The values of $\sqrt{F}$
are chosen to make the lifetimes equal. The dashed line 
shows the parameterization of the range using Eq. (24-25).}
\label{fig:staurange1}
\end{figure}

\begin{figure}[t]
\epsfig{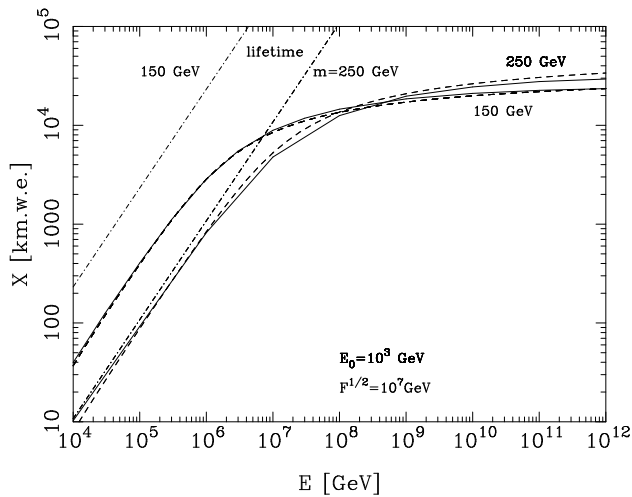}
\caption{Stau range (solid) and time dilated decay length
(dot-dashed line) for
$\mtil=150$ GeV and $\mtil=250$ GeV for $\sqrt{F}=10^7$ GeV.  
The dashed line  shows the parameterization of the range
using Eq. (24-26).}
\label{fig:staurange2}
\end{figure}

The resulting $\beta$ from diagrams (a1) and (a2) of Fig.~\ref{fey1}, 
yielding Eq. (\ref{eq:gval}), 
is obtained for the scalar case and found
to be the same as for the fermion case to within the
numerical accuracy of our integration using the fortran program
VEGAS \cite{lepage} (here, a relative error of $10^{-4})$. 
This can be understood explicitly from Eq. (18), where each factor of
C, which differs for scalar and fermion cases, is multiplied by
$\lambda$. The quantity $\beta$ is dominated by
small $y$, so $\lambda\ll 1$. On a more qualitative level,
one notes that the Gordon decomposition of the fermion current
has a term proportional to $(p_1+p_2)_\mu$, the scalar coupling, which dominates
in the high energy limit.

Because the two photon ((a1) and (a2) diagrams) process 
gives $\beta^{\rm pair}$,
and at high energies, the difference between fermion and scalar
scattering to produce the $e^+e^-$ pair is negligible, 
in the results
we show below we have used the complete fermion
expression of
Ref. \cite{koko}
with the substitution $m_\mu\rightarrow \mtil$ to describe
$\beta^{\rm pair}$ for staus at high energy.

The results for $\beta^{\rm pair}$ are shown by the labeled curve in
Fig.~\ref{fig:beta1} 
for $\mtil=150$ GeV. The scaling with mass is shown in Fig.~\ref{fig:beta2},
where we show $\beta^{\rm pair}$ with solid lines for $\mtil = 50,\
150$
and $250$ GeV. Again, the $1/\mtil$ behavior of $\beta$ appears.
As noted in Ref. \cite{tannenbaum},  the differential cross section
(in $y$ and $Q^2$) can be approximated in the small $y$ limit by a
form similar to Eq. (\ref{eq:fapprox}), 
where instead $f_2\rightarrow \sigma_{e^+
  e^-}$, the photopair production cross section which is nearly
constant,
and $m_0\rightarrow 4m_e$ 
related to the threshold for
electron-positron
pair production. The same Jacobian factor, now $\sim m_e/\mtil$
determines
the $\mtil$ dependence of $\beta^{\rm pair}$.

\subsection{Stau Energy Loss due to Bremsstrahlung}

The bremsstrahlung contribution to $\beta$ is much smaller
than the pair production and photonuclear contributions by
an additional factor of $1/\mtil$. A textbook discussion of
$\beta^{\rm brem}$ \cite{nachtmann} shows that
\begin{equation}
\frac{d \sigma}{dy}\propto \frac{\bar{\sigma}}{y}
\end{equation}
where $\bar{\sigma}=Z^2\alpha^3/\mtil^2$. The expression for
$\beta^{\rm brem}$ therefore has a dependence on the particle mass
through $\bar{\sigma}$, namely $1/\mtil^2$. This is explicitly shown
in Fig.~\ref{fig:beta1} for $\mtil=150$ GeV, where the evaluation of
$\beta$ follows from Ref. \cite{petru}.

\section{Stau Range}
\label{sec:range}

Our evaluation of the stau range follows the same procedure
for tau propagation outlined in Ref. \cite{drss}. Briefly,
we perform a one-dimensional Monte Carlo evaluation which treats
the first two terms in the energy loss
\begin{equation}
-{dE\over dX}=\alpha + {N\over A}E\int_0^{y_{cut}}dy\, y{d\sigma\over dy}
+ {N\over A}E\int^1_{y_{cut}}dy\, y{d\sigma\over dy}\ ,
\end{equation}
continuously, and the third term stochastically. We use
$y_{cut}=10^{-3}$. We evaluate the survival probability 
$P(E,E_0,z)$ as 
a function of initial energy $E$, minimum final energy $E_0$ and
depth $z$. The average range is
\begin{equation}
X(E,E_0)=\int dz\, P(E,E_0,z)\ .
\end{equation}
We note that the average range is not the same as the range of average
energy loss, the solution to Eq. (\ref{eq:dedx}) \cite{ls}. 
Discussed below is a
parameterization of an effective $\beta$ that, when substituted into
the solution to Eq. (\ref{eq:dedx}), gives the approximate average range.

\begin{figure}[t]
\epsfig{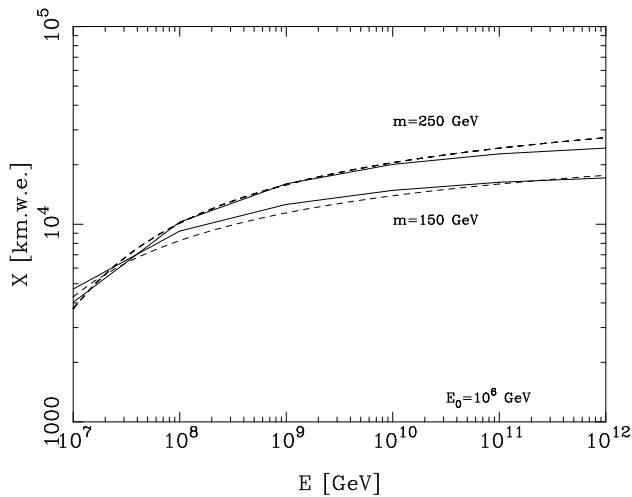}
\caption{Stau range (solid) for $\sqrt{F}=10^7$ GeV and for
$\mtil=150$ GeV and $\mtil=250$ GeV, with $E_0=10^6$ GeV. 
The dashed lines show the parameterization of the range
following Eqs. (24-26) and the solid lines show the Monte
Carlo results.}
\label{fig:staurange3}
\end{figure}

The average ranges of the $\tilde{\tau}$ for two masses, $\mtil=150$ GeV
and $\mtil=250$ GeV are shown in Fig.~\ref{fig:staurange1} 
by the solid lines. For this
figure and Fig.~\ref{fig:staurange2}, 
the stau is propagated until its energy is
reduced to $E_0=10^3$ GeV.
Shown in Fig. \ref{fig:staurange1} 
are the results for $\sqrt{F}=10^7$ GeV for $\mtil=150$ GeV
and $\sqrt{F}=2.15\times 10^7$ GeV for the larger mass. The two
values of $F$ are chosen to give the same lifetimes for $\mtil=150$
and
250 GeV. The time dilated decay length is long compared to the
distance
over which energy is lost by the stau. This is clear in that we
can parameterize the range without reference to the decay length. A
convenient choice is to use the {\it solution} to Eq. (\ref{eq:dedx}) for the
range of average loss with a constant
$\beta$, but to use in the solution an energy dependent $\beta$.
In terms of initial stau energy $E$ and final stau energy $E_0$, we
write
\begin{eqnarray}
X(E,E_0)&=& \frac{1}{\beta} \ln \Bigl[ (\alpha +\beta E)/(\alpha
+ \beta E_0)\Bigr]
\label{eq:range}\\
\beta &\simeq & \beta_0 +\beta_1\ln (E/10^{10}\ {\rm GeV})
\label{eq:betaparam}
\nonumber \\
&& +\beta_2\ln (E_0/10^3\ {\rm GeV}) \ 
\end{eqnarray} 
with
\begin{eqnarray*}
\alpha &=& 2\times 10^{-3}{\ \rm GeV\, cm^2/g}\\
\beta_0 &=& 5\times 10^{-9}{\ \rm  cm^2/g}(150\ {\rm GeV}/\mtil)\\
\beta_1 &=& 2.8\times 10^{-10}{\ \rm  cm^2/g}(150\ {\rm GeV}/\mtil)\\
\beta_2&=& 2\times 10^{-10}{\ \rm  cm^2/g}(150\ {\rm GeV}/\mtil)\ .
\end{eqnarray*}
These parameters are obtained from a fit to the average
range determined by the one-dimensional
Monte Carlo program.
In Fig.~\ref{fig:staurange1}, 
the dashed lines following the solid curves show the
parameterized fit, Eqs. (\ref{eq:range}) and
(\ref{eq:betaparam}), with $E_0=10^3$ GeV.

The average range of stau varies roughly linearly with stau energy $E$ for 
lower energy $E \le 10^6$ GeV in Fig. \ref{fig:staurange1}.  
For $m_{\tilde{l}}=150$ GeV and 
$\sqrt{F}=10^7$ GeV, the range is about 40 km water equivalence (w.e.)
for $E=10^4$ GeV, and about $3\times 10^3$ km w.e. for $E=10^6$ GeV. 
This behavior can be obtained from Eq.~(\ref{eq:range}) where for 
$\beta E \ll \alpha$, $X \propto (E-E_0)/\alpha$. 
For higher energy, the average range varies logarithmically with 
energy: $X \propto \log(E/E_0)/\beta$,  
roughly a few $\times 10^4$ km w.e.

The average range at low energy has little mass dependence since 
$\alpha$ is a nearly a constant as a function of mass.  At 
higher energy, the larger the mass of the particle, the longer 
the average range due to the $1/m_{\tilde{l}}$ dependence of $\beta$.

For $\sqrt{F}$ fixed at $\sqrt{F}=10^7$ GeV, in the case of
$\mtil=150 $ GeV, the low energy range is dominated by ionization
energy loss, while for $\mtil=250$ GeV, it is dominated by the
decay length (see Fig.~\ref{fig:staurange2}).
For the latter case, 
the average range at low energy still has the linear dependence on 
$E$ since $\gamma c \tau = (E/m_{\tilde{l}}) c \tau$.
%The range is shorter than the average range due to the energy loss via
%ionization, since  the stau decays before it loses the energy via ionization.
%
A modification of
Eq. (\ref{eq:range}) to account for the lifetime
involves the substitution
\begin{equation}
\alpha\rightarrow \alpha_c=\alpha + \frac{\mtil}{c\tau\rho}\ .
\end{equation}
With Eq. (\ref{eq:range}) and the energy dependent $\beta$,
the substitution of $\alpha\rightarrow\alpha_c$ leads to a 
reasonable approximation to the range, as seen in Fig.~\ref{fig:staurange2}.
Numerically, for the three range curves shown in 
Figs.~\ref{fig:staurange1} and \ref{fig:staurange2}, the
parameterization agrees to within $\sim 20$\%  over the whole energy
range. In fact, for the curves in Fig.~\ref{fig:staurange1}, 
the ratio of parametrization
of the range to the Monte Carlo evaluation of the range deviates
from 1 only by $-8\% - +1\%$ from $10^4-10^{12}$ GeV initial stau
energy. For the shorter lifetime shown in Fig.~\ref{fig:staurange2}, 
the ratio of the
parameterization to the Monte Carlo range deviates from 1 by
$-5\%- +15\%$ except for $10^4-10^5$ GeV, where the deviation reaches
as much as $-22\%$.

At higher energy thresholds $E_0$, the effective
$\beta$ increases, requiring the $\beta_2$ term in Eq. (\ref{eq:betaparam}).
In Fig.~\ref{fig:staurange3},  
we show the stau range and its parameterization for
$E_0=10^6$ GeV.  The average range is shorter than the 
case of $E_0=10^3$ GeV due to the higher energy threshold. 
The ratio of the parameterization of the range
to the Monte Carlo evaluation goes between $1.03-0.88$ for $\mtil
= 150$ GeV, and between $1.11-0.93$ for $\mtil=250$ GeV.

\begin{figure}[t]
\epsfig{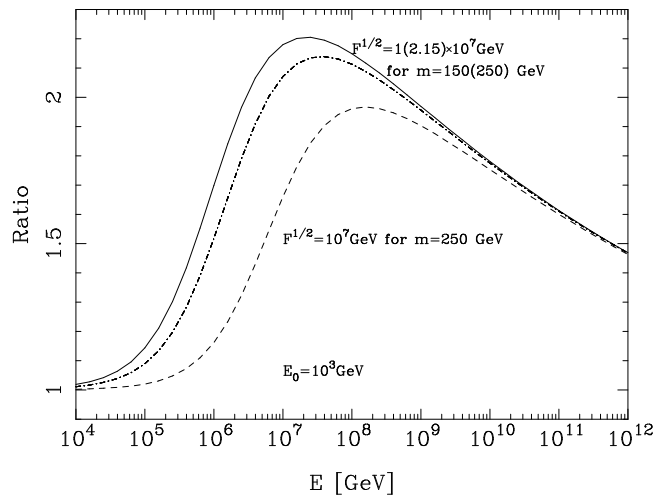}
\caption{Ratio of
stau range using the parameterization of Eqs. (24-26) to stau range
using a constant $\beta=0.8\times 10^{-6} m_\tau/\mtil$ cm$^2$/g
for $\sqrt{F}=10^7$ GeV and $\mtil=150$ GeV 
(solid) and $\mtil=250$ GeV (dashed line), and for 
$\sqrt{F}=2.15\times 10^7$ GeV with $\mtil=250$ GeV 
(dot-dashed line), for $E_0=10^3$ GeV.}
\label{fig:comparison}
\end{figure}

\section{Discussion}
\label{sec:discussion}

Albuquerque, Chacko and Burdman proposed to look for evidence
of supersymmetry in a class of models by looking for pair-produced
staus with nearly parallel tracks \cite{chacko}.
They concluded that there is a potential sensitivity of neutrino
telescopes to a range of parameter space in certain low energy SUSY 
breaking  models. Their
evaluation of
event rates of upward-going staus depends in part 
on the effective area of the detector and the range
of the staus. 

To first approximation, the energy loss of
the stau can be characterized by constant $\alpha$ and
$\beta_{\tilde{\tau},c}$. This is what was done in Ref. \cite{chacko},
where $\beta_\tau\simeq 0.8\times 10^{-6}$ cm$^2$/g for the tau
is rescaled to $\beta_{\tilde{\tau},c}\simeq
9.5\times 10^{-9}\, (150\GeV /\mtil)$ cm$^2$/g according to the
factor of $m_\tau/\mtil$.
With our new estimate of the stau range, parameterized
by an energy dependent $\beta$, the effective range is increased
relative to the range calculated using $\beta_{\tilde{\tau},c}$.

Fig.~\ref{fig:comparison} shows the
ratio of the range using the energy dependent $\beta$ compared to the
range calculated using the constant $\beta_{\tilde{\tau},c}$, 
scaled appropriately
for the different masses, for initial stau energies between $10^4-10^{12}$ GeV
and a minimum stau energy of $E_0=10^3$ GeV. 
The ratio
can be as much as a factor of two for initial stau energies in the
range of $10^7-10^8$ GeV. In this energy range, the $\beta$ term
in Eq. (\ref{eq:dedx}) dominates over the ionization term ($\alpha$) and the
lifetime, but the effective $\beta$ is not as large as the 
rescaled constant $\beta_{\tilde{\tau},c}$ in
this energy range.  
The enhancement of the range
by a factor of two is significant because it substantially improves
the potential of neutrino telescopes
for detecting evidence of stau production.

For $E_0=10^6$ GeV, a similar comparison between the Monte Carlo
range and the range using $\beta_{\tilde{\tau},c}$ is obtained.
The ratio of the range determined by
Monte Carlo propagation to the range evaluated with
Eq. (\ref{eq:dedx}) using $\beta_{\tilde{\tau},c}$
is a factor of $\sim 2$ for $E=10^7-10^8$ GeV for $\mtil=150$ GeV
and a factor of about 1.6 for the same energy range with $\mtil=250$
GeV for $\sqrt{F}=10^7$ GeV.

Estimates which scale the energy loss $\beta$ by $1/\mtil^2$ \cite{bi}
dramatically overestimate the stau range at high energies. Using the
$\beta $ of Ref. \cite{bi}, $\beta=3.9\times 10^{-6} (m_\mu/\mtil)^2$
cm$^2$/g, one finds that for an initial stau energy of $10^6$ GeV,
one has overestimated the range by about 40\%, and at $10^7$ GeV by a
factor
of 4.8. The overestimate rapidly increases 
with energy. It is a factor of $\sim 30$ for
an
initial energy of $10^9$ GeV, a factor of $\sim 1600$ for an initial
energy of $10^{12}$ GeV.

In summary, for the dominant photonuclear and $e^+e^-$ pair production
processes, 
one expects qualitatively that the $\beta$ of a
charged scalar particle (stau)
scales as the inverse mass of the
scalar particle. We have shown this result quantitatively by explicit
calculation. Incorporating the electromagnetic energy loss of the
stau and the long lifetime following from the supersymmetry
breaking scale, we used a Monte Carlo to evaluate the range. 
The range determined by Monte Carlo evaluation is longer
than the range of average energy loss using Eq. (\ref{eq:dedx}) and a constant
$\beta_{\tilde{\tau},c}$. 
Our energy dependent parametrization of an effective $\beta$ in 
Eq. (\ref{eq:betaparam})
for the stau range reasonably approximates
the Monte Carlo results. Our evaluation of the stau range suggests 
that previous estimates \cite{chacko} of upward stau pair event
rates,
induced by  supersymmetric
particle production by upward-going neutrinos, may be enhanced
by a factor of $\sim 2$ in some energy ranges.

\acknowledgments

This work was supported in part by DOE contracts DE-FG02-91ER40664,
DE-FG02-04ER41319 and DE-FG02-04ER41298 (Task C).
We thank F. Halzen, S. Brodsky and M. Tannenbaum
for helpful conversations.

\end{document}